\documentclass{article}
\usepackage{graphics}
\usepackage{latexsym}
\usepackage{amsmath}
\usepackage{amsthm}
\usepackage{amssymb}
\usepackage{amsfonts}
\usepackage{amsxtra}
\usepackage{hyperref}
\allowdisplaybreaks
\title{Light pseudo-Goldstone bosons without explicit symmetry breaking}
\author{Antonio O.\ Bouzas \thanks{E-mail:
    abouzas@mda.cinvestav.mx}\\\small Departamento de F\'{\i}sica
  Aplicada, CINVESTAV-IPN \\\small Carretera Antigua a Progreso Km.\
  6, Apdo.\ Postal 73 ``Cordemex''\\\small
  M\'erida 97310, Yucat\'an, M\'exico}
\date{September, 2006}
\newcommand{\sss}[1]{{\scriptscriptstyle #1}}
\newcommand{\mat}[1]{\ensuremath{\boldsymbol{#1}}}

\newcommand{\Ll}{\ensuremath{\mathcal{L}}}

\newcommand{\naught}{\sss{0}}
\newcommand{\dirac}{\not\negmedspace\partial}

\newcommand{\tr}{\mathrm{Tr}}
\newcommand{\Tr}[1]{\mathrm{Tr}\left( #1 \right)}
\newcommand{\diag}{\mathrm{diag}}
\newcommand{\rank}{\mathrm{rank}}

\newcommand{\xo}{{x^{\naught}}}
\newcommand{\yo}{{y^{\naught}}}
\newcommand{\Yo}{{\mathcal{Y}^{\naught}}}

\newcommand{\wV}{\widetilde{V}}
\newcommand{\dv}[2]{#1^{(#2)}}
\newcommand{\SUN}[1]{\ensuremath{\mathrm{SU}(#1)}}
\newcommand{\sun}[1]{\ensuremath{\mathrm{su}(#1)}}
\newcommand{\G}{\ensuremath{\mathrm{G}}}
\newcommand{\tomega}{\widetilde{\omega}}
\newcommand{\ts}{t_\mathrm{s}}
\newcommand{\tf}{t_\mathrm{f}}
\begin{document}
\maketitle
\begin{abstract}
  A mechanism is discussed to obtain light scalar fields from a
  spontaneously broken continuous symmetry without explicitly breaking
  it.  If there is a continuous manifold of classical vacua in orbit
  space, its tangent directions describe classically massless fields
  that may acquire mass from perturbations of the potential that do
  not break the symmetry.  We consider the simplest possible example,
  involving a scalar field in the adjoint representation of SU(N).  We
  study the scalar mass spectrum and its RG running at one-loop level
  including scalar and pseudoscalar Yukawa couplings to a massive
  Dirac fermion.
\end{abstract}
\section{Introduction}
\label{sec:intro}

In a theory with a global or local symmetry group \G\ spontaneously
broken by a multiplet of scalar fields, the classical vacuum manifold
$\mathcal{Y}$ is globally invariant under the action of \G.  The set
$\mathcal{Y}/\G$ of classical minima in orbit space is often
discrete.  When $\mathcal{Y}/\G$ is continuous its tangent directions
are null vectors of the second derivative of the potential and,
therefore, describe classically massless modes which are not Goldstone
bosons (henceforth referred to as GBs) associated to the spontaneous
breaking of \G.  It may happen that these tangent modes are GBs of a
larger symmetry group $\mathrm{G}'\supset \G$ of the scalar sector,
explicitly broken by other interaction terms invariant under \G\ but
not $\mathrm{G}'$.  In other words, they may be pseudo-Goldstone
bosons (henceforth PGBs) related to $\mathrm{G}'$.  If no such larger
symmetry $\mathrm{G}'$ exists, the tangent modes may acquire mass from
additional small \G-invariant terms in the scalar potential which
cause deformations of $\mathcal{Y}$.  We have then light scalar modes
without explicitly breaking the symmetries of the theory, neither at
tree level nor radiatively.

In this paper we consider the simplest possible model of light PGBs
without explicit symmetry breaking, involving a scalar field in the
adjoint representation of \SUN{N}, $N>3$.  We study the mass spectrum
of the model, and its renormalization when the scalar fields couple to
Dirac fermions.  
In dimensional regularization fermion loops may have a large effect on
the renormalization group (henceforth RG) running of scalar couplings
and masses.  In particular, as discussed below, fermionic radiative
corrections may cause the mass of tangent modes to be decreasing
functions of the renormalization scale.  Because our aim is to
investigate the mechanism outlined above to obtain light scalar
fields, for simplicity we omit gauge interactions.

In section \ref{sec:vacu} below we briefly review the extremum
analysis for a quartic potential in the adjoint representation of
\SUN{N}.  In section \ref{sec:massp} we consider the classical mass
spectrum in the vacua of interest to us.  The renormalization group
evolution of that spectrum is studied in section \ref{sec:RG}, where
we also consider in detail the conditions for stability of the vacuum.
We try to keep the overlap of these sections with the previous
literature, particularly \cite{xli74,mur83}, to a minimum.  Final
remarks are given in section \ref{sec:fin}, and some related material
is gathered in two appendices.

\section{Classical vacua}
\label{sec:vacu}

We consider a scalar field $\phi(x)$ in the adjoint representation of
\SUN{N}, $N\geq 3$.  The most general renormalizable potential
$V_N:\sun{N} \rightarrow \mathbb{R}$ invariant under $\SUN{N}$ is,
\begin{equation}
  \label{eq:pot}
  V_N(\phi) = -\frac{m^2}{2}\Tr{\phi^2} + c \Tr{\phi^3} + a
  (\Tr{\phi^2})^2 + b \Tr{\phi^4}~. 
\end{equation}
We restrict ourselves to the strictly renormalizable case, $a\neq 0
\neq b$, with spontaneous symmetry breaking, $m^2>0$.  We introduce
the reduced potential $V_N(x) \equiv V_N(\diag(x_1,\ldots,x_N))$, with
$\sum_{j=1}^N x_j = 0$.  Ignoring the tracelessness constraint for the
moment, the condition for $V_N(x)$ to be bounded below is that
$V_N^{(4)}(x) \equiv a (\sum_{j=1}^N x_j^2 )^2 + b \sum_{j=1}^N x_j^4
> 0$ for all $x$.  Since $V_N^{(4)}(x)$ is an homogeneous function of
$x$ of even degree, this is equivalent to requiring that the minimum
value of $V_N^{(4)}(x)$ over the unit sphere be positive.  That
minimum value is $a+b$ if $a>0$ and $Na+b$ if $a<0$, so a necessary
and sufficient condition for $V_N$ to be bounded below is
\begin{equation}
  \label{eq:pot1}
  a+b>0 \quad \text{and} \quad Na+b>0~.
\end{equation}
When the constraint $\sum_{j=1}^N x_j = 0$ is taken into account the
conditions (\ref{eq:pot1}) are only sufficient, not necessary.
We will always assume, nevertheless, that $a$ and $b$
are chosen so that (\ref{eq:pot1}) are satisfied.

To find the extrema of $V_N(x)$ we consider the unconstrained
extremization of $\Ll_N(x,\lambda) = V_N(x_1,\ldots,x_N) +
\lambda\sum_{j=1}^N x_j = 0$, with $\lambda$ a Lagrange multiplier.
$\Ll_N(x,\lambda)$ is a quartic symmetric polynomial in $x_i$.  If
$\xo$ is an extremum of $\Ll_N$, $N\geq 4$, there cannot be more than
three different values among its components if $b\neq 0$.  To see
this, we assume otherwise.  Since $\Ll_N$ is invariant under
permutations of $x_i$, we can always assume that $\xo$ is such that
$\xo_1<\xo_2<\xo_3<\xo_4$.  Thus, $\xo$ is a point interior to the
open domain $\{x_1<x_2<x_3<x_4\}\subset\mathbb{R}^N$, and within that
domain we can use as coordinates $s_1, s_2, s_3, s_4, x_5,\ldots,
x_N$, with $s_k=\sum_{j=1}^4 x_j^k$.  A necessary condition for $\xo$
to be an extremum is then $(\partial\Ll_N/\partial s_4)|_\xo=b=0$.
Therefore, for $b\neq 0$ all extrema of the potential must be either
of the form,
\begin{subequations}
  \label{eq:pot3}
\begin{gather}
  \label{eq:pot3a}
  \xo=(\underbrace{\eta_1,\ldots,\eta_1}_{n_1},
  \underbrace{\eta_2,\ldots,\eta_2}_{n_2},
  \underbrace{\eta_3,\ldots,\eta_3}_{n_3})~,
  \quad
  n_1+n_2+n_3=N~,
  \quad
  n_{1,2,3}\geq 1\\
  \label{eq:pot3b}
  n_1 \eta_1 + n_2 \eta_2 + n_3 \eta_3 = 0~,\\
  \intertext{or of the form}
  \xo=(\underbrace{\eta_1,\ldots,\eta_1}_{k_1},
  \underbrace{\eta_2,\ldots,\eta_2}_{k_2})~,
  \quad
  k_1+k_2=N~,
  \quad
  k_{1,2}\geq 1~,
  \quad
  \label{eq:pot3c}
  k_1 \eta_1 + k_2 \eta_2=0~.
\end{gather}
\end{subequations}
Since $V_N$ is symmetric in its arguments $x_i$, any permutation of
the components of $\xo$ is also an extremum.  We can then adopt the
convention $n_1\geq n_2 \geq n_3$. In some exceptional cases discussed
below, an extremum can be both of type (\ref{eq:pot3a}) (with, say,
$\eta_1=\eta_2$) and of type (\ref{eq:pot3c}). 

In the case of extrema $\xo$ of the form (\ref{eq:pot3a}) with
$\eta_i\neq \eta_j$ if $i\neq j$ ($i,j=1,2,3$), we can argue as above
to conclude that $\eta_i$ must satisfy $(\partial\Ll_N/\partial
s_3)|_\xo = (c+4/3 b s_1)|_\xo=0$ or, more explicitly,
\begin{equation}
  \label{eq:pot4}
  \eta_1+\eta_2+\eta_3 = -\frac{3}{4} \frac{c}{b}~.
\end{equation}
Taking into account (\ref{eq:pot3a}), (\ref{eq:pot3b}), and
(\ref{eq:pot4}), the extremum equations $\partial\Ll_N/\partial
x_i =0$ can be reduced to
\begin{equation}
  \label{eq:pot5}
  (2n_1 a +b) \eta_1^2 + (2 n_2 a +b) \eta_2^2 + (2 n_3 a +b) \eta_3^2
  - \frac{m^2}{2} - \frac{9 c^2}{16 b} = 0~.
\end{equation}
The constrained extrema of $V_N(x)$ are then either of the form
(\ref{eq:pot3a}), with $\eta_{1,2,3}$ determined by (\ref{eq:pot3b}),
(\ref{eq:pot4}) and (\ref{eq:pot5}), or of the form (\ref{eq:pot3c}).
The latter can be obtained by setting $n_3= 0$ in the solutions
of type (\ref{eq:pot3a}), with eqs.\ (\ref{eq:pot3b}) and
(\ref{eq:pot5}) determining $\eta_{1,2}$ in terms of $\eta_3$ as
defined by (\ref{eq:pot4}).  In order to find the extrema of $V_N(x)$
we consider the case $c=0$ first, and the general case afterwards.
Since $V_N(x)$ is invariant under the transformation $x\rightarrow
-x$, $c\rightarrow -c$, we can restrict ourselves to $c\geq 0$~.

\paragraph{Even potential  ($\mat{c=0}$)}

The solutions to eqs.\ (\ref{eq:pot3b}), (\ref{eq:pot4}),
(\ref{eq:pot5}) in this case are, with notation analogous to
(\ref{eq:pot3a}),
\begin{subequations}
\label{eq:solu1}
\begin{gather}
  \yo_{n_1,n_2,n_3}=(\underbrace{\xi_1,\ldots,\xi_1}_{n_1},
  \underbrace{\xi_2,\ldots,\xi_2}_{n_2},
  \underbrace{\xi_3,\ldots,\xi_3}_{n_3})~,
  \quad
  n_1+n_2+n_3=N~,
  \quad
  n_{1,2,3}\geq 1  \label{eq:solu1a}\\
  \label{eq:solu1b}
  \xi_1=\frac{m}{2}\frac{n_2-n_3}{\sqrt{\Omega_{n_1,n_2,n_3}}}~,
  \quad
  \xi_2=\frac{m}{2}\frac{n_3-n_1}{\sqrt{\Omega_{n_1,n_2,n_3}}}~,
  \quad
  \xi_3=\frac{m}{2}\frac{n_1-n_2}{\sqrt{\Omega_{n_1,n_2,n_3}}}~,\\ 
  \begin{gathered}
  \Omega_{n_1,n_2,n_3} = \Omega^{n_1,n_2,n_3}_a a +
  \Omega^{n_1,n_2,n_3}_b b,
  \quad
  \Omega^{n_1,n_2,n_3}_a = N (n_1 n_2 +n_2 n_3 +n_3 n_1) - 9 n_1 n_2 n_3,\\
  \Omega^{n_1,n_2,n_3}_b = N^2 - 3  (n_1 n_2 +n_2 n_3 +n_3 n_1)~.    
  \end{gathered}
  \label{eq:solu1c} 
\end{gather}
\end{subequations}
Besides $\pm \yo_{n_1,n_2,n_3}$, all permutations of their components
are also solutions to the extremum eqs.  Clearly, if $n_1=n_2=n_3$
(\ref{eq:solu1}) reduces to the trivial solution.  If not all $n_i$
are equal it is possible to prove that $0<\Omega^{n_1,n_2,n_3}_a$,
$0<\Omega^{n_1,n_2,n_3}_b$, and $1\leq
\Omega^{n_1,n_2,n_3}_a/\Omega^{n_1,n_2,n_3}_b \leq N$, provided
(\ref{eq:pot1}) holds.  Therefore $\Omega_{n_1,n_2,n_3}>0$, and these
extrema exist.

If $N=3 n$ it is possible to have $n_1=n_2=n_3=n$.  In this case eqs.\
(\ref{eq:pot3b}) and (\ref{eq:pot4}) are not independent so, instead
of a discrete set, the extrema lie on the curve $\Yo_{n,n,n}$ defined
by
\begin{equation}
  \label{eq:solu2}
  \xi_1 +   \xi_2 + \xi_3 = 0~,
  \qquad
  \xi_1^2 +   \xi_2^2 +   \xi_3^2 = \kappa_n^2~,
  \qquad   
  \left(\kappa_n^2 \equiv \frac{m^2}{2(2n a+b)}\right)~.
\end{equation}
If (\ref{eq:pot1}) holds, then $(2na+b)>0$.  For $N=3$ these are
minima. 

We can find all two-valued extrema $\yo_{n_1,n_2}$ directly by setting
$n_3=0$ in (\ref{eq:solu1}), $\yo_{n_1,n_2} \equiv \yo_{n_1,n_2,0}$.
Using (\ref{eq:solu1}) it can be shown that there cannot be integers
$n_{1,2,3}\geq 1$ and $k_{1,2}\geq 1$ with $n_1+n_2+n_3=N=k_1+k_2$
such that $\yo_{n_1,n_2,n_3}$ and $\yo_{k_1,k_2}$ are equal, up to a
permutation of their components.  If $N=3n$, however, some of the
$\Yo_{n,n,n}$ may have only two different values among their
components.  In fact, $\yo_{2n,n}$ with components $\xi_{1,2}$ as
given by (\ref{eq:solu1b}) with $n_3=0$ are easily seen to satisfy
(\ref{eq:solu2}), with $\xi'_1=\xi_1$, $\xi'_2=\xi_1$, $\xi'_3=\xi_2$.
Thus, $\pm \yo_{2n,n}$ and their permutations are part of the set
$\Yo_{n,n,n}$.

\paragraph{Non-even potential ($\mat{c>0}$)}

The solutions to the non-homogeneous linear eqs.\ (\ref{eq:pot3b}) and
(\ref{eq:pot4}) can be written as $\eta_i = \sigma\xi_i + \rho_i$,
$i=1,2,3$, with $\xi_i$ the solution to the homogeneous eqs.\ given by
(\ref{eq:solu1}), $\sigma$ a free real parameter, and $\rho_i$ a
particular solution to (\ref{eq:pot3b}) and (\ref{eq:pot4}). In the
case $n_1>n_2>n_3\geq 1$, $\rho_i$ can be taken to be,
\begin{equation}
  \label{eq:solurho}
  \rho_1 = \frac{3c}{4b}\frac{n_2 n_3}{n_{12}n_{31}}~,
  \quad
  \rho_2 = \frac{3c}{4b}\frac{n_1 n_3}{n_{12}n_{23}}~,
  \quad
  \rho_3 = \frac{3c}{4b}\frac{n_1 n_2}{n_{23}n_{31}}~,
  \quad
   n_{ij} \equiv n_i-n_j~,
\end{equation}
and similarly in the cases $n_2=n_3$ and $n_1=n_2$.  There are no
solutions with $n_1=n_2=n_3$. The parameter $\sigma$ is determined
from (\ref{eq:pot5}), which reduces to
\begin{equation}
  \label{eq:solus}
  \frac{m^2}{4}\sigma^2 + \sum_{i=1}^3 (2 a n_i+b) \xi_i\rho_i \sigma
  + \frac{1}{2} \sum_{j=1}^3 (2 a n_j+b) \rho_j^2 = \frac{m^2}{4} +
  \frac{9 c^2}{32 b} ~.
\end{equation}
The extrema of type (\ref{eq:pot3a}) are then explicitly determined by
(\ref{eq:solu1}), (\ref{eq:solurho}), and the two solutions to
(\ref{eq:solus}).  There are also extrema of the form (\ref{eq:pot3c})
wich are given by (\ref{eq:solu1}) and (\ref{eq:solurho}), with
$n_3=0$, and
\begin{equation}
  \label{eq:solusi}
  \sigma^2 - 2 z \sigma -1=0~,
  \qquad
  z=\frac{3c}{4m}\frac{n_1-n_2}{\sqrt{\Omega_{n_1,n_2,0}}}~.
\end{equation}
We will denote the two possible values of $\sigma$ by $\sigma_\pm = z
\pm \sqrt{z^2+1}$, and the corresponding extrema as
$\xo_{n_1,n_2,\pm}$.  Since in this case $\rho_{1,2}=0$, we have
$\xo_{n_1,n_2,\pm} = \sigma_\pm \yo_{n_1,n_2}$.  

Provided the conditions (\ref{eq:pot1}) are satisfied, for fixed
values of $a$, $b\neq0$, $m^2>0$, and $n_{1,2,3}\geq 1$, there exists
at least one value of $c^2$ such that two of the $\eta_i$ in
(\ref{eq:pot3a}) are equal.  For those values of $c$ the extremum
$\xo$ can be written both in the form (\ref{eq:pot3a}) (with $n_3\geq
1$) and (\ref{eq:pot3c}).  We will not dwell longer on this issue
because it will not be of importance in what follows and, furthermore,
because such fine tuning of parameters is not preserved under
renormalization.

\section{Mass spectrum}
\label{sec:massp}

At each extremum $\xo$ the Hessian matrix for the potential $V_N(x)$
can be diagonalized (see appendix \ref{sec:hess}) to obtain the
corresponding mass spectrum.  In the case $c=0$ this diagonalization
can be carried out explicitly and the extrema $\yo_{n_1,n_2,n_3}$ and
$\yo_{n_1,n_2}$ completely classified.  When $c\neq 0$ a complete
classification is more difficult to obtain by direct computation.  A 
proof that all extrema $\xo_{n_1,n_2,n_3}$ with three different
components are saddle points is given in \cite{mur83}.
At an extremum $\xo_{n_1,n_2,\pm}$ ($n_1\geq n_2\geq 1$) the Hessian
matrix has the four eigenvalues,  
\begin{equation}
  \label{eq:masspec}
\begin{aligned}
  \omega_1 &= 0~, &\qquad & [1] &&\{2n_1 n_2+1\}\\
  \omega_2 &= m^2(\sigma^2_\pm+1)~, &\qquad  &[1]&&\{1\}\\
  \omega_3 &= m^2 N \left(\frac{2 z \sigma_\pm}{n_1-n_2} + \frac{2
      n_2-n_1}{\Omega_{n_1,n_2,0}}\sigma_\pm^2 b\right)~,  &\qquad
  &[n_1-1] &&\{n_1^2-1\}\\ 
  \omega_4 &= m^2 N \left(-\frac{2 z \sigma_\pm}{n_1-n_2} + \frac{2
      n_1-n_2}{\Omega_{n_1,n_2,0}}\sigma_\pm^2 b\right)~,  &\qquad  &
  [n_2-1] && \{n_2^2-1\}~. 
\end{aligned}
\end{equation} 
The square brackets indicate ``reduced'' multiplicities, corresponding
to the Hessian matrix of the reduced potential $V_N(x)$.  These are
multiplicities in orbit space (see appendix \ref{sec:appb}).  The
``total'' multiplicities obtained from the Hessian matrix of the full
potential $V_N(\phi)$ on $\sun{N}$, indicated in (\ref{eq:masspec}) in
curly brackets, can be computed from reduced multiplicities (appendix
\ref{sec:appb}), or simply by noticing that the stability group of
$\xo_{n_1,n_2,\pm}$ is $\SUN{n_1}\times \SUN{n_2}\times \mathrm{U}(1)$
if $n_2>1$, and $\SUN{n_1}\times \mathrm{U}(1)$ if $n_2=1$.  One null
eigenvalue ($\omega_1$ in (\ref{eq:masspec})) is an artifact of the
projection on the constrained subspace $\sum_{i=1}^N x_i=0$ (see
appendix \ref{sec:hess}).  The eigenvalues $\omega_{2,3,4}$ give the
squared masses of three massive multiplets. The mass-squared average
and difference between the $\SUN{n_1}$ and $\SUN{n_2}$ multiplets
is\footnote{The average mass in (\ref{eq:masspec2}) is slightly
  different from the result in \cite{mur83}.}
\begin{equation}
  \label{eq:masspec2}
  \frac{1}{2}(\omega_3+\omega_4)= \frac{m^2}{2}\frac{N^2
    b}{\Omega_{n_1,n_2,0}} \sigma_\pm^2~,
  \qquad
  \omega_3 - \omega_4 = m^2 N \left( \frac{4 z \sigma_\pm}{n_1-n_2} -
    3 \frac{n_1-n_2}{\Omega_{n_1,n_2,0}} b \sigma_\pm^2\right)~.
\end{equation}
Notice that $\omega_3+\omega_4 < 0$ if $b<0$ and $n_2>1$, so
$\xo_{n_1,n_2,\pm}$ cannot be minima in that case .  In some open
regions of the plane $(a,b)$ it is possible to choose $c=c(a,b)$ so 
that $\omega_3 - \omega_4=0$ but, with the exception of the case
$n_1=n_2$, $c=0$, such fine tuning is not preserved under
renormalization.

For even potentials ($c=0$) if $N=3n$ there is an extremal manifold
$\Yo_{n,n,n}$ defined by (\ref{eq:solu2}), containing in particular
the extrema $\pm\yo_{2n,n}$.  We are interested in the mass spectrum
at those extrema and, for $c\neq 0$, at the related extrema
$\xo_{2n,n,\pm}$.  We consider the case $N=3n$ with $n>1$ first.  The
mass-squared spectrum of $V_N(x)|_{c=0}$ at the extrema satisfying
(\ref{eq:solu2}) and such that $\xi_i\neq \xi_j$ if $i\neq j$ is (see
appendix \ref{sec:hess})
\begin{equation}
  \label{eq:contspec}
  \begin{gathered}
\tomega_1 = 0~,\quad [2]\{6n^2+2\}~,
\qquad
\tomega_2 = 2 m^2~,\quad [1]\{1\}~,
\\
\tomega_3 = 12 b \left(\xi_1^2 - \frac{\kappa_n^2}{6}\right)~,\quad 
\tomega_4 = 12 b \left(\xi_2^2 - \frac{\kappa_n^2}{6}\right)~,\quad
\tomega_5 = 12 b \left(\xi_3^2 - \frac{\kappa_n^2}{6}\right),\quad
[n-1]\{n^2-1\}~,
  \end{gathered}
\end{equation}
where we used the same notation as in (\ref{eq:masspec}) and
$\kappa_n^2$ is defined in (\ref{eq:solu2}). All the eigenvalues on
the last line have the same multiplicity.  The null eigenvalue
$\tomega_1$ has two eigenvectors in orbit space, $\sun{N}/\SUN{N}$, as
indicated in (\ref{eq:contspec}).  One of them is a spurious
projection mode.  The other one lies on the subspace tangent to the
1-dimensional manifold defined by (\ref{eq:solu2}), on which
$V_N(x)|_{c=0}$ is constant.  The stability group of these extrema is
$\SUN{n}^3\times \mathrm{U}(1)^2$, so we expect to have $6n^2$ GBs
plus one spurious massless mode, three massive $\SUN{n}$ multiplets,
and two independent massive $\mathrm{U}(1)$ modes.  Out of the latter,
one lies in the tangent subspace and remains massless at tree level.
Since it is impossible to satisfy (\ref{eq:solu2}) and to have
$(\xi_i^2-\kappa_n^2/6) >0$ (or $<0$) simultaneously for $i=1,2,3$,
these extrema are saddle points for any values of $a$ and $b$.

The extrema of $V_N(x)|_{c=0}$ satisfying (\ref{eq:solu2}) with
$\xi_1=\xi_2$ are $\pm\yo_{2n,n}$ and their permutations, with
stability group $\SUN{2n}\times \SUN{n}\times \mathrm{U}(1)$.  The
eigenvalue spectrum at $\yo_{2n,n}$ can be obtained either from
(\ref{eq:masspec}) (with $c=0$, $n_1=2n_2=2n$) or from
(\ref{eq:contspec}) (with $\xi_1=\xi_2=\pm\kappa_n/\sqrt{6}$,
$\xi_3=-2\xi_1$).  The result is that there is a null eigenvalue with
total multiplicity $8n^2$ comprising $4 n^2$ GBs, one spurious
projection mode, and $4n^2-1$ PGBs.  The latter form an $\SUN{2n}$
multiplet containing the mode lying in the tangent direction to the
manifold (\ref{eq:solu2}) in orbit space.  Since the tangent mode is
massless at tree level, the mass of the entire multiplet must vanish.
The spectrum contains also a $\mathrm{U}(1)$ mode of squared mass $2
m^2$ and an $\SUN{n}$ multiplet with $n^2-1$ modes of mass
$6b\kappa_n^2$.  We see that for $b>0$ all Hessian eigenvalues at
$\pm\yo_{2n,n}$ are non-negative.  Expanding $V_N(x)|_{c=0}$ in power
series about $\pm\yo_{2n,n}$ through third order, however, shows that
these extrema are not minima but saddle points.

When $c\neq 0$ there are no solutions to (\ref{eq:solu2}).  The only
extrema related to the extremal manifold $\Yo_{n,n,n}$ are
$\xo_{2n,n,\pm}$ as defined in \S \ref{sec:vacu}, with $\xo_{2n,n,\pm}
\xrightarrow{c\rightarrow 0} \pm\yo_{2n,n}$. As discussed above, we
need only consider the case $c>0$.  In this case it is immediate from
(\ref{eq:masspec}) that $\xo_{2n,n,-}$ cannot be a minimum, because at
it $\omega_3<0$.  On the other hand, at $\xo_{2n,n,+}$ the non-zero
eigenvalues in the spectrum (\ref{eq:masspec}) reduce to
\begin{equation}
  \label{eq:masspecito}
\begin{gathered}
  \omega_2=m^2 (1+\sigma_{+}^2) = 2 m^2 +
  \mathcal{O}\left(cm\right)~, 
  \qquad
  \omega_3 = 6 m^2 z \sigma_{+} = \frac{3\sqrt{3}}{2}
  \frac{cm}{\sqrt{2 n a+b}} +
  \mathcal{O}\left(c^2\right)~,\\
  \omega_4 = 6 \kappa_n^2 (2 n a (1-\sigma_+^2)+b) = 6 \kappa_n^2 b
  + \mathcal{O}\left(cm\right)~, 
\end{gathered}
\end{equation}
with $\kappa_n^2$ defined in (\ref{eq:solu2}) and $z$ and $\sigma_+$ in
(\ref{eq:solusi}).  The counting of modes is the same as in the case
of $\yo_{2n,n}$ discussed above, except that in (\ref{eq:masspecito})
the $\SUN{2n}$ multiplet has a mass $\sqrt{\omega_3} =
\mathcal{O}(\sqrt{cm})$.  The eigenvalue $\omega_2$ is obviously
positive and, since we are assuming $c>0$ and (\ref{eq:pot1}), also
$\omega_3>0$.  The conditions on the coupling constants for
$\omega_4>0$, and for $\omega_4\lessgtr \omega_3$, can be summarized
as follows
\begin{align}\label{eq:monster}
&a>0~,\quad b>0  \quad \text{and}\quad 
    \begin{cases}
       0<\frac{c}{m} < \sqrt{\frac{2}{3}}\frac{b}{\sqrt{4na+b}} 
      & \Rightarrow  \omega_4 > \omega_3 > 0~,\\
       \sqrt{\frac{2}{3}}\frac{b}{\sqrt{4na+b}} < \frac{c}{m} <
      \sqrt{\frac{2}{3na}}b & \Rightarrow  \omega_3 >
      \omega_4 >
      0~,\\
      \sqrt{\frac{2}{3na}}b < \frac{c}{m} & \Rightarrow 
      \omega_3 > 0 > \omega_4~,
    \end{cases}\nonumber\\
&a>0~,\quad b<0  \Rightarrow   \omega_3 > 0 >  \omega_4 ~,
\\
&a<0~,\quad b>0 \quad \text{and}\quad
  \begin{cases}
    4na+b < 0  & \Rightarrow  \omega_4 > \omega_3 > 0~,\\
    4na+b > 0 \quad\text{and}\quad \frac{c}{m} <
    \sqrt{\frac{2}{3}} \frac{b}{\sqrt{4na+b}}
    & \Rightarrow 
    \omega_4 > \omega_3 > 0~,\\
    4na+b > 0 \quad\text{and}\quad
    \sqrt{\frac{2}{3}}\frac{b}{\sqrt{4na+b}} < \frac{c}{m} 
    & \Rightarrow  \omega_3 > \omega_4 > 0~.
  \end{cases}\nonumber
\end{align}
We remark that (\ref{eq:monster}) holds under the assumptions $N=3n$,
$n>1$, $c>0$, and (\ref{eq:pot1}).

In the case $N=3$ ($n=1$), the eigenvalues of the Hessian matrix of
$V_3(x)|_{c=0}$ at the extremal manifold $\Yo_{1,1,1}$ defined by
(\ref{eq:solu2}) are $\tomega_1=0$ (with reduced and total
multiplicities $[2]\{8\}$, resp.), and $\tomega_2=2 m^2$ (with
$[1]\{1\}$).  One of the null eigenvectors in orbit space is a
spurious projection mode, and the other one lies on the subspace
tangent to $\Yo_{1,1,1}$, on which $V_3(x)|_{c=0}$ is constant.  Since
there are no further eigenvalues, these extrema are minima
independently of the values of $a$ and $b$, as long as the potential
remains bounded below.  $V_3(\phi)|_{c=0}$ is invariant under
$\mathrm{SO}(8)$ (because $\tr{\phi^4}=1/2(\tr{\phi^2})^2$ in
$\sun{3}$), spontaneously broken to $\mathrm{SO}(7)$.  Thus, the 7
massless modes belonging to $\tomega_1$ are GBs arising from this
spontaneous symmetry breaking.  At the extrema $\pm\yo_{2,1}$ the
$\SUN{3}$ subgroup of stability is $\SUN{2}\times \mathrm{U}(1)$, so
there are 4 GBs, a massive $\mathrm{U}(1)$ mode and an $\SUN{2}$
triplet of PGBs, which can acquire mass from $\mathrm{SO}(8)$ breaking
interactions.  This triplet contains the tangent mode.  When $c\neq 0$
the spectrum at $\xo_{2,1,+}$ is as in (\ref{eq:masspecito}), with
$n=1$, and with $\omega_4$ omitted.  The triplet of PGBs acquires a
mass of $\mathcal{O}(\sqrt{cm})$ from the explicit breaking of
$\mathrm{SO}(8)$ due to $c\neq 0$.

Unlike the case $N=3$, for $N>3$ the symmetry $\mathrm{SO}(N^2-1)$ is
explicitly broken by the dimension 4 operator $\tr(\phi^4)$ in
$V_N(\phi)$ so we expect it to play a limited role in the theory, and
not to have any influence on the scalar mass spectrum for $|b|\gtrsim
|a|$. 

\section{Renormalization group evolution}
\label{sec:RG}

We consider now the RG evolution of the masses and the parameters in
the potential.  We consider the interaction of the scalar field $\phi$
with a fermion field in the fundamental representation of $\SUN{N}$,
\begin{equation}
  \label{eq:lagrangian}
  \Ll = \frac{1}{4}\tr\left(\partial_\mu\phi\partial^\mu\phi\right) -
  V_N(\phi) +\overline{\psi} i\dirac\psi- \overline{\psi} (M+i M_5
  \gamma_5) \psi -\overline{\psi}\phi (g+ig_5\gamma_5)\psi~.
\end{equation}
The one-loop RG equations in MS scheme for the Yukawa couplings and
fermion masses are, 
\begin{subequations}
  \label{eq:fermi}
\begin{align}
  &\begin{aligned}
    \mu\frac{dg}{d\mu} &\equiv\beta_g =\frac{1}{8\pi^2N}
    (N^2-3)g g_5^2 +\frac{1}{8\pi^2N}(N-1)(N+3) g^3
    \\
    \mu\frac{dg_5}{d\mu} &\equiv\beta_{g_5}
    =\frac{1}{8\pi^2N}(N-1)(N+3) g_5 g^2 + \frac{1}{8\pi^2N} 
    (N^2-3) g_5^3
  \end{aligned}\label{eq:fermia}\\
  &\begin{aligned}
    \mu\frac{dM}{d\mu} &= \frac{1}{8\pi^2N} (N^2-1)
    \left( \rule{0pt}{11pt} M (3g^2-g_5^2) + 4 gg_5 M_5\right)   \\ 
    \mu\frac{dM_5}{d\mu} &= \frac{1}{8\pi^2N} (N^2-1)\left(
      \rule{0pt}{11pt} M_5 (3g_5^2-g^2) + 4 gg_5 M\right)~.
    \label{eq:fermib}
  \end{aligned}
\end{align}
\end{subequations}
For the dimensionless parameters in $V_N$ we have, 
\begin{subequations}
  \label{eq:scab}
\begin{align}
  \label{eq:scaba}
   \mu\frac{da}{d\mu} &\equiv\beta_a = \frac{2}{\pi^2}(N^2+7) a^2 +
   \frac{6}{\pi^2N^2} (N^2+3) b^2+\frac{4}{\pi^2N}(2N^2-3) ab
   +\frac{1}{\pi^2} a g^2\\ 
   \mu\frac{db}{d\mu} &\equiv\beta_b = \frac{4}{\pi^2N}(N^2-9) b^2 + 
   \frac{24}{\pi^2} a b +\frac{1}{\pi^2}bg^2 -\frac{1}{8\pi^2}
   (g^2+g_5^2)^2~,   \label{eq:scabb}
\end{align}
\end{subequations}
and for the dimensionful ones,
\begin{subequations}
\label{eq:dimful}
\begin{align}
   \mu\frac{dc}{d\mu} &= \frac{6}{\pi^2}c \left(
     \left(N-\frac{6}{N}\right) b + 2 
     a +\frac{1}{8} g^2 \right) - \frac{1}{2\pi^2} (Mg+M_5
   g_5)(g^2+g_5^2)\label{eq:dimfula}\\
   \mu \frac{dm^2}{d\mu} &= \frac{2}{\pi^2} m^2 \left((N^2+1)
     a+\frac{2N^2-3}{N} b + \frac{g^2}{4}\right) - \frac{9}{\pi^2N}
   (N^2-4) c^2 \label{eq:dimfulb}\\
   &\quad + \frac{1}{\pi^2}(g^2+g_5^2)(M^2+M_5^2) +
   \frac{2}{\pi^2}(gM+g_5M_5)^2~.\nonumber  
\end{align}
\end{subequations}
Some comments about these equations are in order.  $\beta_g$ and
$\beta_{g_5}$ have the same sign as $g$ and $g_5$, resp., so $|g|$ and
$|g_5|$ are monotonically increasing functions of $\mu$, as expected.
The coupling $a$ is monotonically increasing (see below), but $|a|$
and $b$ need not be and, therefore, the stability of the potential is
not guaranteed, as discussed in detail below.  The sign of the fermion
contribution to $\beta_b$ is independent of $N$, and of whether the
scalar field is in the adjoint representation of $\SUN{N}$ or
$\mathrm{U}(N)$. In the case $N=1$ our sign agrees with textbook
results (see ch.\ 5 of \cite{col93} and ch.\ 6 of \cite{che91}).  If
$M=0=M_5$, then $\mu dc/d\mu\propto c$, due to chiral symmetry.  Eqs.\
(\ref{eq:fermia}), (\ref{eq:scab}) and (\ref{eq:dimfulb}) agree with
\cite{bou03} when $g_5=0=M=M_5=c=0$.\footnote{Except for the terms of
  $\mathcal{O}(g^4)$, which were considered of higher order in
  \cite{bou03}. Here we take into account the full one-loop
  contribution.} An important verification is that for $N=3$ the
quantities $\beta_a+1/2 \beta_b$, $\mu dc/d\mu$ and $\mu dm^2/d\mu$
should not depend on $a$ and $b$ separately, but only on $a+b/2$.
This test is passed by (\ref{eq:scab}) and
(\ref{eq:dimful}). \footnote{But failed by the RG eqs.\ in
  \cite{mur83}.}  

The running Yukawa couplings are easily obtained by setting $g(t)=g_i
f(t)$, $g_5(t)=g_{5i}f(t)$, $f(0)=1$, with $g_i$, $g_{5i}$ the initial
values, $f(t)$ an auxiliary function and $t\equiv \log(\mu/\mu_0)$,
with $\mu\geqslant\mu_0>0$ and $\mu_0$ a reference $\mathrm{MS}$
scale.  Solving (\ref{eq:fermia}) for $f(t)$ we get
\begin{equation}
  \label{eq:yuk}
  g(t)=\frac{g_i}{\sqrt{1-t/\tf}}~,
  \quad
  g_5(t)=\frac{g_{5i}}{\sqrt{1-t/\tf}}~,
  \quad
  \tf=\frac{4\pi^2 N}{(N^2-3) g_{5i}^2+(N-1)(N+3)g_i^2}~.
\end{equation}
The fact that $g$ and $g_5$ enter asymmetrically in $\beta_g$ and
$\beta_{g5}$, and therefore also in $\tf$, is due to the scalar wave
function renormalization not depending on $g_5$ at this order.  From
(\ref{eq:yuk}) we see that the ratio $g/g_5$ is RG invariant. With
(\ref{eq:yuk}), (\ref{eq:fermib}) can also be integrated
\begin{equation}
  \label{eq:ferm}
\begin{aligned}
  M(t) &= \frac{g_i}{g_i^2+g_{5i}^2} (g_i M_i+g_{5i}M_{5i}) \left(
  \frac{1}{1-t/\tf}\right)^{\gamma_1} + 
  \frac{g_{5i}}{g_i^2+g_{5i}^2} (g_{5i} M_i-g_{i}M_{5i}) \left(
    \frac{1}{1-t/\tf}\right)^{\gamma_2}~,\\
  M_5(t) &= \frac{g_{5i}}{g_i^2+g_{5i}^2} (g_i M_i+g_{5i}M_{5i}) \left(
  \frac{1}{1-t/\tf}\right)^{\gamma_1} - 
  \frac{g_{i}}{g_i^2+g_{5i}^2} (g_{5i} M_i-g_{i}M_{5i}) \left(
    \frac{1}{1-t/\tf}\right)^{\gamma_2}~,\\
  \gamma_1&=-3\gamma_2 = \frac{3}{8\pi^2}\frac{N^2-1}{N} \tf
  (g_i^2+g_{5i}^2) ~.
\end{aligned}
\end{equation}
We remark that $\gamma_1=3/2+\mathcal{O}(N^{-2})$ and
$\gamma_2=-1/2+\mathcal{O}(N^{-2})$, independently of the asymptotic
behavior of Yukawa couplings for large $N$.  A combination with a
particularly simple analytical expression is $g(t) M(t)+g_5(t) M_5(t)
= (g_i M_i+g_{5i} M_{5i}) (1/(1-t/\tf))^{\gamma_1+1/2}$.  Clearly, the
one-loop approximation involved in (\ref{eq:fermi}) only holds for
$|g|, |g_5|\ll 1$, so the exact solutions (\ref{eq:yuk}) and (\ref{eq:ferm})
are physically valid only for $t\ll \tf$.

We consider now (\ref{eq:scaba}).  For $a$ and $g^2$ fixed $\beta_a$
is a quadratic polynomial in $b$ and $\beta_a<0$ is possible only if
$b$ lies between its roots.  From the expression for those roots we
can show that $\beta_a<0$ implies that (\ref{eq:pot1}) does not hold.
Thus, if $V_N$ is bounded below then $\beta_a>0$ and $a(t)$ is
monotonically increasing.  Like $g(t)$ and $g_5(t)$, $a(t)$ and $b(t)$
also have a mobile singularity at a finite $t$, whose location $\ts>0$
we can choose as a constant of integration.  (There is a related
singularity for $t<0$ which will not be of interest to us). For
$g^2+g_5^2$ large enough the evolution of $a$ and $b$ is dominated by
those couplings and $\ts=\tf$.  We assume $g^2, g_5^2 \sim
\mathcal{O}(|a|)$ or smaller.  Under that assumption we have $\ts<\tf$
and, in fact, over large regions of parameter space $\ts\ll \tf$.

In the region near the singularity $b(t)$ can be neglected in
(\ref{eq:scaba}) and, with (\ref{eq:yuk}), it can be integrated to
give 
\begin{equation}
  \label{eq:approxa}
\begin{aligned}
  a(t) &\simeq \frac{\pi^2}{2(N^2+7)} \frac{A}{(\tf-t)\left(
      1-\left(\frac{\tf-\ts}{\tf-t}\right)^A\right)} \simeq
  \frac{\pi^2}{2(N^2+7)} \frac{1}{\ts-t}~,\\
  A &= \frac{(N+1)(N-3)g_i^2+(N^2-3)g_{5i}^2}{(N-1)(N+3)
    g_i^2+(N^2-3)g_{5i}^2}~.
\end{aligned}
\end{equation}
The second equality on the first line holds for $t\ll \tf$.  If
$g_i=0$ we must set $\tf=\infty$ in (\ref{eq:approxa}). For
$|b_i|<a_i$ the expression (\ref{eq:approxa}) gives a good
approximation to $a(t)$ also in the physically relevant region $t\ll
\ts$.  Furthermore, for $N>3$ we see that $A\simeq 1$ and from the
initial condition $a(0)=a_i$ we obtain \footnote{From (\ref{eq:yuk})
  and (\ref{eq:time}), $\ts<\tf \Longleftrightarrow (N^2-3) g_{5i}^2 +
  (N-1)(N+3) g_i^2 < 8\pi^2 N (N^2+7) a_i$.  Notice that the rhs is
  cubic in $N$ and the lhs is quadratic.  With $8\pi^2 \simeq 80$ and
  $g_i^2$, $g_{5i}^2 \simeq \mathcal{O}(a_i)$, the inequality is
  satisfied in large open regions of parameter space.  }
\begin{equation}
  \label{eq:time}
  \ts\simeq \frac{\pi^2}{2(N^2+7) a_i}~.
\end{equation}
If $|b_i|>|a_i|$ the approximate solution (\ref{eq:approxa}) does not
hold.  There is a critical value $t_1$, however, such that
$a(t)>|b(t)|$ for all $t_1<t<\ts$.  Thus, for $|b_i|>|a_i|$ we can
find an approximate solution that interpolates between the regime
$|b(t)|>|a(t)|$ ($0\leq t < t_1$) and the regime $|b(t)|<a(t)$ where
(\ref{eq:approxa}) holds.  

We are interested in the UV stability of the potential and of the
vacuum.  By UV stability of the potential we mean that (\ref{eq:pot1})
is satisfied for all $0<t<\ts$ if it is satisfied at $t=0$.
Similarly, we say that the vacuum is UV stable if it is a minimum of
$V_N$ for $0\leq t < \ts$.  In the discussion of UV stability that
follows the relevant regime is $|b(t)|<a(t)$ so we assume $|b_i|<a_i$
for that purpose (but not in the numerical solutions given below).

Stability of the vacuum requires $b>0$ (see (\ref{eq:monster})).
However, the condition $b_i>0$ is not enough to guarantee that
$b(t)>0$ for all $0<t<\ts$, since the last term in (\ref{eq:scabb})
can drive $b(t)$ to negative values.  Once $b$ becomes negative it
diverges to $-\infty$ at $t=\ts$ driven by the last three terms in
(\ref{eq:scabb}).  There is a minimal value $b_\mathrm{min}>0$ such
that $b(t)>0$ for all $0<t<\ts$ if and only if $b_i>b_\mathrm{min}$.
In order to obtain an approximate expression for $b_\mathrm{min}$ the
relevant regime is $b<a$, so we can neglect the first term on the rhs
of (\ref{eq:scabb}) and integrate the resulting linear eq.\ with
$a(t)$ from (\ref{eq:approxa}) and $g$, $g_5$ from (\ref{eq:yuk}), to
get,
\begin{equation}
  \label{eq:approxb}
  b(t)\simeq
  \left(\frac{1}{1-t/\ts}\right)^{12/(N^2+7)}
  \left( e^{g_i^2 t/\pi^2} \left( b_i - \frac{1}{8}
      \frac{(g_i^2+g_{5i}^2)^2}{g_i^2} \right) +  
      \frac{1}{8} \frac{(g_i^2+g_{5i}^2)^2}{g_i^2}\right)~.
\end{equation}
This approximate solution, together with (\ref{eq:time}), leads to 
\begin{equation}
  \label{eq:bmin}
  b_\mathrm{min} \simeq \frac{1}{8} \frac{(g_i^2+g_{5i}^2)^2}{g_i^2}
  \left( 1 - e^{-g_i^2/(2(N^2+7)a_i)}\right)~.
\end{equation}
The condition $b_i>b_\mathrm{min}$ is not very restrictive since,
for $g_i^2, g_{5i}^2 \sim \mathcal{O}(a_i)$, typically
$b_\mathrm{min} \ll a_i$.  The accuracy of the approximate relation
(\ref{eq:bmin}) is best illustrated by quoting some representative
numerical values.  For $a_i=1/600$, $g_i=1/10=g_{5i}$,
(\ref{eq:bmin}) gives a result for $b_\mathrm{min}$ that differs
from the result obtained numerically from (\ref{eq:scabb}) by 35\% for
$N=3$, 10\% for $N=6$ and 3\% for $N=12$.  The accuracy is higher for
larger values of $N$.  

Since we require $b>0$ for $0\leq t<\ts$, and since $a$ is
monotonically increasing, if $a_i>0$ then (\ref{eq:pot1}) are
trivially satisfied for all $0\leq t <\ts$ and $V_N$ is UV stable.  If
$a_i<0$, but such that (\ref{eq:pot1}) are satisfied at $t=0$ with
$b_i>b_\mathrm{min}$, (\ref{eq:pot1}) can in principle be violated at
$t>0$ if $b(t)>0$ decreases rapidly enough.  This imposes bounds on
the possible values of $a_i<0$, $g_i$ and $g_{5i}$, which must be
satisfied in order to ensure that $Na+b$ remains positive throughout
the evolution.  Since, however, $a(t)$ is monotonically increasing and
$b(t)>0$, (\ref{eq:pot1}) will be satisfied for all values of $t$
$(<\ts)$ larger than a certain critical value.  We will not dwell
longer on this, and from now on we assume $a_i>0$,
$b_i>b_\mathrm{min}>0$ so that $a(t), b(t)>0$ for all $0\leq t<\ts$.

We assume $V_N$ to be of the form (\ref{eq:pot}) with $c\geq0$, the
case $c<0$ can be obtained by means of the transformation
$\phi\rightarrow -\phi$.  The fermion masses $M$, $M_5$ and Yukawa
couplings $g$, $g_5$ can be $\lessgtr 0$, their signs entering the
scalar sector RG eqs.\ (\ref{eq:scab}), (\ref{eq:dimful}) only through
$Mg+M_5 g_5$ in (\ref{eq:dimfula}).  With the assumptions $a_i>0$,
$b_i>b_\mathrm{min}$, and $c_i>0$, if $Mg+M_5g_5<0$ then $dc/dt>0$ and
$c(t)$ is monotonically increasing, therefore positive throughout the
evolution.  If $Mg+M_5g_5>0$, however, even if $c_i>0$ it may happen
that $c(t)$ becomes negative, thus diverging to $-\infty$ at $t=\ts$.
The classical extremum $\xo_{2n,n,+}$ is a saddle-point for $c<0$.  As
in the case of $b(t)$ discussed above, we will have $c(t)>0$ for all
$0\leq t<\ts$ only if $c_i>c_\mathrm{min}$ for some critical value
$c_\mathrm{min}$.  An approximate solution to (\ref{eq:dimfula}),
valid for $b_i<a_i$, can be obtained by substituting
(\ref{eq:approxa}), (\ref{eq:approxb}) (with $\exp(g_i^2 t/\pi^2)\simeq
1$), (\ref{eq:yuk}) and (\ref{eq:ferm}) into (\ref{eq:dimfula}) to
obtain,
\begin{equation}
  \label{eq:approxc}
  \begin{gathered}
  c(t)\simeq
  \left( \frac{1}{1-t/\ts}\right)^{6/(N^2+7)}
  \left( e^{\rho t}(c_i-\widetilde{c}_i) + \widetilde{c}_i)\right)~,\\
  \rho=\frac{3}{4\pi^2} g_i^2 + \frac{6}{\pi^2} \frac{N^2-6}{N}
  \frac{N^2+7}{N^2-5} b_i~,
  \quad
  \widetilde{c}_i = \frac{1}{2\pi^2\rho}
  (M_ig_i+M_{5i}g_{5i})(g_i^2+g_{5i}^2)~. 
  \end{gathered}
\end{equation}
From here, we get,
\begin{equation}
  \label{eq:cmin}
  c_\mathrm{min}\simeq
  \left(1- e^{-\rho \ts}\right) \widetilde{c}_i~.
\end{equation}
Notice that for $Mg+M_5 g_5<0$ we have $c_i>0>c_\mathrm{min}$, so
$c(t)>0$ for all $0<t<\ts$. Substituting (\ref{eq:time}) in
(\ref{eq:cmin}) we obtain an approximate expression for
$c_\mathrm{min}$ solely in terms of initial values.  To give an
estimate of the accuracy of that approximation, for $a_i=1/300$,
$b_i=1/450$, $g_i=-1/15=g_{5i}$, $M_i=M_{5i}=m_i=1$, the value of
$c_\mathrm{min}$ obtained with (\ref{eq:cmin}) and (\ref{eq:time})
differs from the more accurate value computed numerically from
(\ref{eq:dimfula}) by 35\%, 23\% and 11\% for $N=3$, 6 and 12, resp.

The evolution of $m^2(t)$ may be dominated by the term $\propto c^2$
in (\ref{eq:dimfulb}), for $c^2(t)$ sufficiently large, causing
$m^2(t)$ to decrease and eventually become
negative.\footnote{Typically, $\omega_4$ changes sign at lower scales
  than $m^2$ does (see below), in principle precluding the possibility
  of radiative symmetry breaking, at least in this vacuum.}  An upper
bound $c_\mathrm{max}$ on the initial value $c_i$ exists, so that
$c_i<c_\mathrm{max}$ ensures that $m^2(t)>0$ for $0\leq t<\ts$.  Using
the exact solutions for the fermion parameters and the approximate
ones for $a$, $b$, $c$ given above, we find the following approximate
form from $m^2(t)$, valid for $0<b_i<a_i$,
\begin{equation}
  \label{eq:approxm}
  \begin{gathered}
    m^2(t)\simeq
    \left( \frac{1}{1-t/\ts}\right)^\frac{N^2+1}{N^2+7}
    \left( m_i^2 e^{\eta t} + \widetilde{m}^2_i t
      \left(1-\frac{1}{2}\frac{t}{\ts}\right) \right)~,
    \qquad
    \eta=\frac{2}{\pi^2}\frac{2N^2-3}{N} b_i+\frac{g_i^2}{2\pi^2}~. \\
    \widetilde{m}^2_i = \frac{1}{\pi^2}
    \left( -9 \frac{N^2-4}{N} c_i^2 +
      (g_i^2+g_{5i}^2)(M_i^2+M_{5i}^2)+2(g_iM_i + g_{5i} M_{5i})^2
    \right)~.
  \end{gathered}
\end{equation}
From this approximate solution we obtain, using (\ref{eq:time}),
\begin{equation}
  \label{eq:cmax}
  c_\mathrm{max} \simeq \frac{1}{3}\sqrt{\frac{N}{(N^2-4)}}
  \left( 4(N^2+7) a_i m_i^2 + (g_i^2+g_{5i}^2)(M_i^2+M_{5i}^2) + 2
    (g_i M_i+g_{5i} M_{5i})^2 \right)^{1/2}~.
\end{equation}
For the set of parameters quoted below (\ref{eq:cmin}) the values for
$c_\mathrm{max}$ obtained from (\ref{eq:cmax}) differ from those
obtained numerically from (\ref{eq:dimfulb}) by 23\%, 9\% and 3\% for
$N=3$, 6 and 12, resp.

A further condition for the stability of the vacuum is $\omega_4>0$.
From (\ref{eq:monster}) we see that such condition can be formulated
as $r(t)<1$, with $r=\sqrt{N/2} a^{1/2} c/(b m)$.  A necessary
condition for $r(t)<1$ throughout the evolution is obviously $r(0)<1$,
\begin{equation}
  \label{eq:obvious}
  \sqrt{\frac{N}{2}} a^{1/2}_i c_i < b_i m_i~.
\end{equation}
The dependence of $r(t)$ on $t$ is not monotonic in general, so
imposing (\ref{eq:obvious}) and $r({\ts}^-)<1$ is not sufficient to
guarantee that $r(t)<1$ for all $t<\ts$.  Furthermore, $r(t)$ may have
rapid variations, which makes it difficult to obtain approximate
expressions for it that remain accurate over broad regions of initial
values in parameter space.  For these reasons, we will not attempt to
give a single bound on initial conditions that ensures $\max_{0\leq
  t<\ts}(r(t))<1$.  Rather, we require initial conditions to obey
(\ref{eq:obvious}), and study the evolution of $\omega_4$ numerically.
The stability condition $\omega_4>0$ implies an upper bound on $c_i$
which is typically more restrictive than (\ref{eq:cmax}).

\begin{figure}
\scalebox{0.5}{\includegraphics{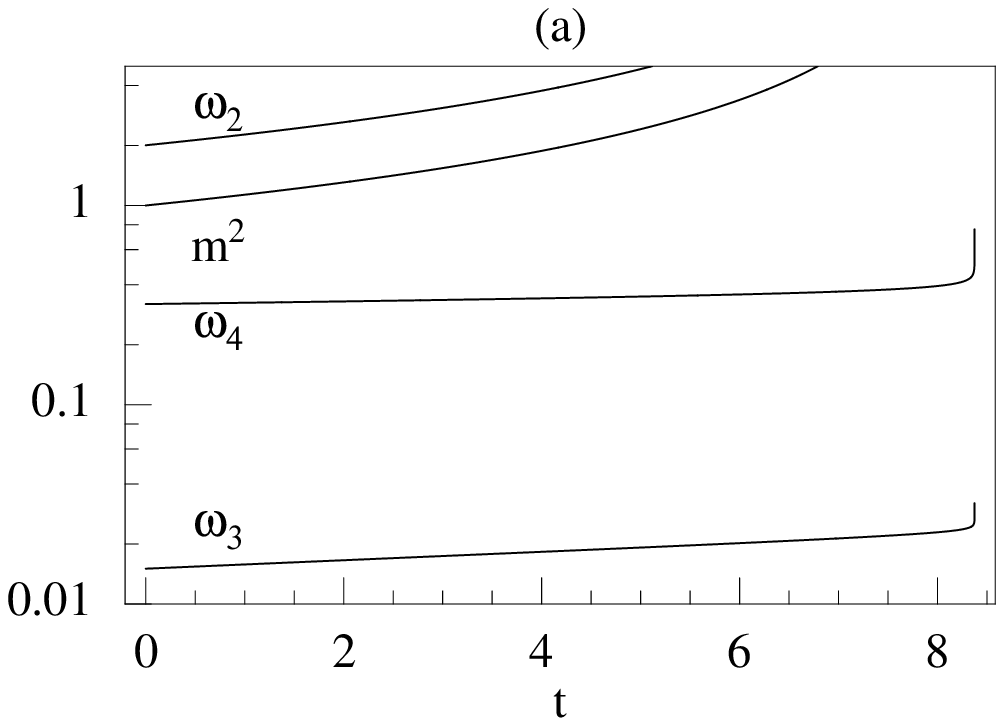}}    
\scalebox{0.49}{\includegraphics{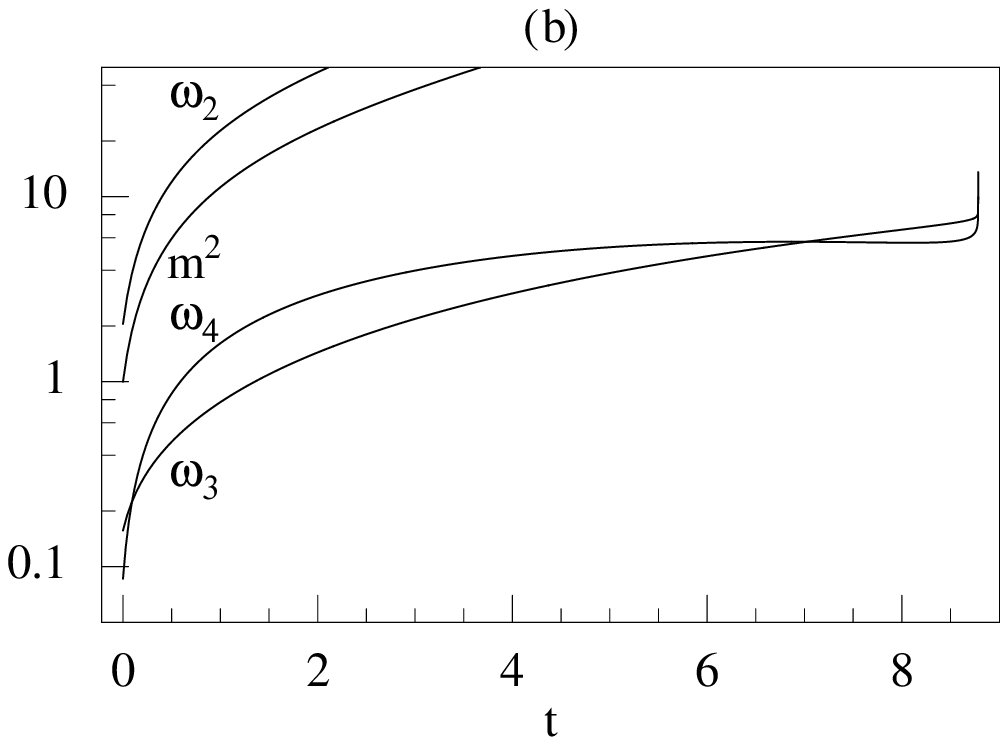}}    
\scalebox{0.5}{\includegraphics{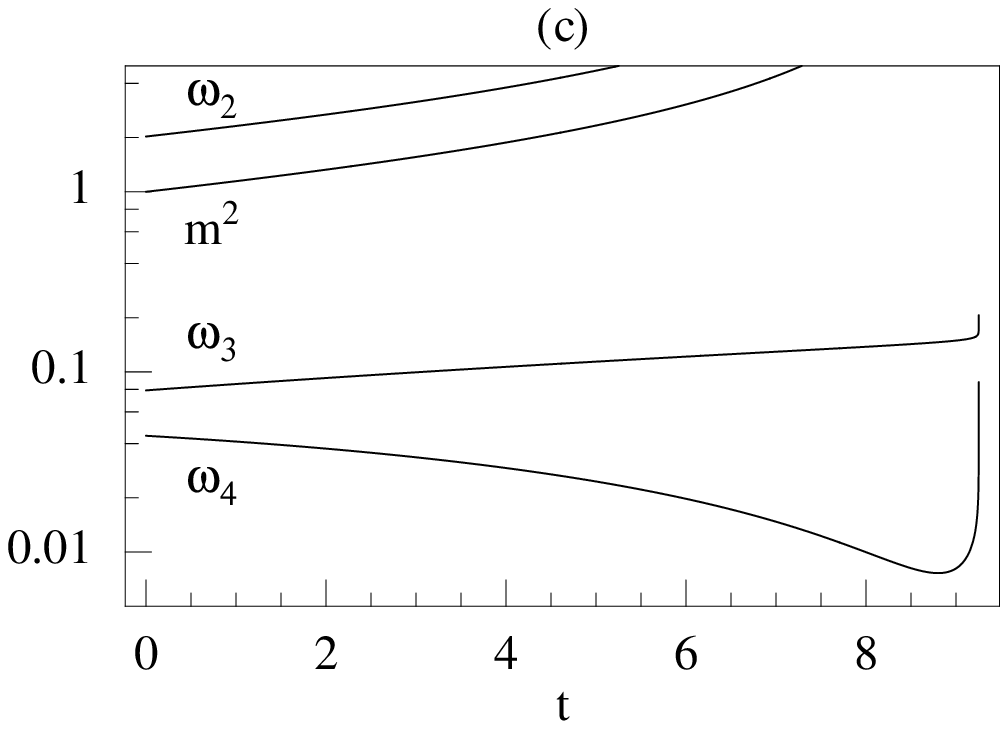}}    
  \caption{RG evolution of $m^2(t)/m_i^2$ and $\omega_{2,3,4}/m_i^2$, for $M
    g+M_5 g_5<0$.  In all cases $N=12$, $a_i=1/300$. 
    (a) $b_i=1/300$, $c_i=1/1000$, $g_i=-1/20=g_{5i}$, $M_i=1=M_{5i}$,
    (b) $b_i=1/450$, $c_i=1/100$, $g_i=-1/10=g_{5i}$, $M_i=20$,
    $M_{5i}=35$,
    (c) $b_i=1/900$, $c_i=1/200$, $g_i=-1/10=g_{5i}$,
    $M_i=3/2=M_{5i}$.
    The end values of $a$ are: (a) $a(8)=0.09$, (b)
    $a(8.5)=0.11$, (c) $a(9)=0.13$.}
  \label{fig:fig}
\end{figure}

Fig.\ \ref{fig:fig} shows the evolution of the mass-squared
eigenvalues (\ref{eq:masspecito}) for $Mg+M_5 g_5<0$.  As shown there,
the location $\ts$ of the RG flow singularity has a small dependence
on $b_i$, not taken into account in (\ref{eq:time}) which is strictly
valid only for $b_i\ll a_i$.  $\omega_{3,4}$ remain much smaller than
$\omega_2$ and $m^2$ throughout the evolution, even for $b\gtrsim a$,
when the broken $\mathrm{SO}(N^2-1)$ symmetry should be irrelevant.

Depending on the values of $b_i$ and $c_i$, $\omega_3$ can be smaller,
larger, or approximately equal to $\omega_4$ throughout the evolution,
as seen in fig.\ \ref{fig:fig} and (\ref{eq:monster}).  The effect of
fermion masses on the evolution of scalar ones is apparent in fig.\
\ref{fig:fig}b.  In that figure the value of $c$ is close to the upper
bound imposed by (\ref{eq:obvious}), which is reflected in the small
value of $\omega_4$ at the beginning of the evolution.  Similarly, a
slightly larger value of $c$ in fig.\ \ref{fig:fig}c would cause the
dip at the end of the evolution of $\omega_4$ to reach negative
values.

\begin{figure}
\centering
\scalebox{0.5}{\includegraphics{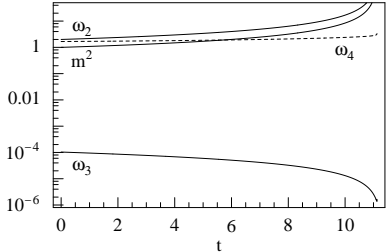}}      
\caption{As in fig.\ \ref{fig:fig}, but for $Mg+M_5 g_5>0$.
  $N=12$, $a_i=1/1000$, $b_i=1/100$, $c_i=5.35 \times 10^{-6}$,
  $g_i=1/30=g_{5i}$, $M_i=1/10=M_{5i}$. The end value of $a$ is 
  $a(10.5)=0.05$.}
  \label{fig:fig2}
\end{figure}

For $Mg+M_5g_5>0$ the RG evolution of the mass-squared spectrum
remains qualitatively the same as in fig.\ \ref{fig:fig}, except for 
the case $c\simeq c_\mathrm{min}$ in which $\omega_3(t)$ (as well as
$c(t)$) is monotonically decreasing over essentially all of the
interval $0\leq t<t_\mathrm{s}$.  This decrease is shown in fig.\
\ref{fig:fig2}, where we chose a small value for $a_i$ and $b_i\simeq
10 a_i$ in order to obtain a larger $\ts$ and a small value of
$c_\mathrm{min}$.  For the parameters used in the figure, with
$c_i\simeq c_\mathrm{min}$, $\omega_3$ is separated from the rest of
the spectrum by a factor $10^4$--$10^6$ throughout the evolution.

\section{Final remarks}
\label{sec:fin}

At the beginning of sect.\ \ref{sec:vacu} we describe the qualitative
motivations for this paper.  The two possibilities mentioned there are
realized in the model defined by (\ref{eq:pot}) and
(\ref{eq:lagrangian}), at the classical vacuum $\xo_{2n,n,+}$ defined
in sect.\ \ref{sec:massp}.  First, in the case $N=3$ the scalar sector
is invariant under $\mathrm{SO}(8)$, spontaneously broken to
$\mathrm{SO}(7)$ and explicitly broken by the fermion sector and also,
softly, by the cubic scalar self-coupling if $c\neq 0$.  That explicit
breaking gives mass to a triplet of PGBs, as described in sect.\
\ref{sec:RG}.  When $c=0$ there is a manifold of minima in orbit
space, $\Yo_{1,1,1}$ described by (\ref{eq:solu2}), which collapses to
the two points $\xo_{2,1,\pm}$ for $c\neq 0$.  Second, in the case
$N=3n>3$ there is no larger symmetry group containing \SUN{N}, and no
explicit symmetry breaking.  When $c=0$ the manifold $\Yo_{n,n,n}$ of
(\ref{eq:solu2}) in orbit space comprises only saddle points, and
collapses into $\xo_{2n,n,\pm}$ for $c>0$.  At the minimum
$\xo_{2n,n,+}$ there is an \SUN{2n} multiplet, containing the mode
tangent to $\Yo_{n,n,n}$ when $c\rightarrow 0$, with mass
$\propto\sqrt{cm}$.  This happens even if (at a given renormalization
scale $\mu$) we set $b\gg |a|\geq 0$ so that, unlike the case $N=3$,
there is no hint of $\mathrm{SO}(N^2-1)$ symmetry in \Ll.  We remark
that, as long as $b\neq 0$, there is no eigenvalue proportional to $c$
in the spectrum (\ref{eq:masspec}) at any extremum $\xo_{n_1,n_2,\pm}$
except for $n_1=2n_2$.  

In section \ref{sec:RG} we discuss in detail the RG running of
couplings and masses.  Approximate bounds are found on the initial
values of $b$ and $c$ which are necessary for UV stability of the
vacuum.  Numerical study of the evolution of $m^2$ and
$\omega_{2,3,4}$ (figs.\ \ref{fig:fig} and \ref{fig:fig2}) shows that
$\omega_3$ (and, for $b\ll a$, also $\omega_4$) remains much smaller
than $m^2$.  For $gM+g_5M_5<0$, $c(t)$ is monotonically increasing but
for $gM+g_5M_5>0$ we have $c>0$ (as needed for $\omega_3>0$) only if
$c(0)>c_\mathrm{min}$, with $c_\mathrm{min}$ approximately given
(\ref{eq:cmin}).  Interestingly, for $c(0)$ close enough to (but still
larger than) its lower bound the mass $\sqrt{\omega_3}$ of the
``tangent mode'' multiplet is a monotonically decreasing function of
the renormalization scale throughout the evolution.

For $N>3$ the vacuum $\xo_{2n,n,+}$ is not a global minimum, which in
the simple model discussed here is not important. In fact, even in
realistic models a metastable vacuum is not in conflict with
phenomenology if long--lived enough.  Although the mass splittings we
obtain with this model are somewhat modest, we may speculate that the
mechanism described in this paper could lead to more elaborate
theories in which the massless modes in the spectrum are removed by
the Higgs mechanism and the heavier ones are pushed up by radiative
corrections to very high mass scales, so that only the lighter massive
modes (corresponding to $\omega_3$ in fig.\ \ref{fig:fig2}) remain in
the physically accesible spectrum.

\appendix

\section{Projected Hessian matrix for constrained extrema}
\label{sec:hess}

Let $V:\mathbb{R}^N\rightarrow\mathbb{R}$ and
$g_a:\mathbb{R}^N\rightarrow\mathbb{R}$, $1\leq a\leq A <N$, be smooth
functions.  Let $\mathcal{G}\subset\mathbb{R}^N$ be the implicit
manifold defined by the set of equations $g_a(x)=0$.  We assume that
$\rank (\partial_j g_a(x)) = A$ for all $x\in\mathcal{G}$,
so that $\mathcal{G}$ is a smooth manifold.  We are interested in the
extrema of $V|_\mathcal{G}:\mathcal{G}\rightarrow\mathbb{R}$.
Let $\xo$ be such an extremum, $c:[-1,1]\rightarrow\mathcal{G}$ a
smooth curve such that $c(0)=\xo$, and $\wV (t) = V(c(t))$.  We must
have $\dot{\wV}(0) =
\partial_i V(\xo)\dot{c}_i(0) = 0$, and since this must be true for
any vector $\dot{c}(0)$ tangent to $\mathcal{G}$ at $\xo$, we conclude
that $\partial_i V(\xo)=\sum_{a=1}^A \lambda_a^\naught \partial_i
g_a(\xo)$ for some set of numbers $\lambda_a^\naught$. Thus, we are
led to an extremization problem for $\Ll(x,\lambda_a) = V(x) -
\sum_{a=1}^A \lambda_a \partial_i g_a(x)$, which is the method of
Lagrange multipliers.

The sign of $\ddot{\wV}(0) = \partial^2_{ij} V(\xo) \dot{c}_i(0)
\dot{c}_j(0) + \partial_i V(\xo) \ddot{c}_i(0)$ determines the nature
of the extremum $\xo$.  Twice differentiating the relation $g(c(t))=0$
and expressing $\partial_i V(\xo)$ in terms of Lagrange multipliers we
get 
\begin{equation}
\label{eq:lagra1}
\ddot{\wV}(0) = \left(\partial^2_{ij} V(\xo) -
  \sum_{a=1}^A\lambda_a^\naught \partial^2_{ij} g(\xo)\right)
  \dot{c}_i(0)\dot{c}_j(0)~.
\end{equation}
For $\xo$ to be a minimum (resp.\ maximum) the expression on the
l.h.s.\ must be positive (negative) for all tangent vectors
$\dot{c}(0)$.  Therefore, the Hessian matrix in which we are
interested is
\begin{equation}
  \label{eq:lagra2}
  \widetilde{H}_V(\xo) = P_\mathcal{G}(\xo)\left(H_V(\xo) -
    \sum_{a=1}^A\lambda_a^\naught H_{g_a}(\xo)\right)P_\mathcal{G}(\xo)~, 
\end{equation}
where $(H_V(\xo))_{ij}=\partial^2_{ij} V(\xo)$ and similarly
$H_{g_a}(\xo)$, and $P_\mathcal{G}(\xo)$ is the projector onto the
subspace tangent to $\mathcal{G}$ at $\xo$.  In the simplest case in
which the constraints $g_a$ are mutually orthogonal ($\partial_k g_a
\partial_k g_b \propto\delta_{ab}$) we have
\begin{equation}
  \label{eq:lagra3}
  (P_\mathcal{G})_{ij}(\xo) = \delta_{ij} - \sum_{a=1}^A
  \frac{1}{\partial_k g_a(\xo) \partial_k g_a(\xo)} \partial_i
  g_a(\xo) \partial_j g_a(\xo)~. 
\end{equation}
Due to the projection $P_\mathcal{G}(\xo)$ the matrix
$\widetilde{H}_V(\xo)$ has $A$ spurious null eigenvectors that span
the subspace normal to $\mathcal{G}$ at $\xo$ and can be taken to be
$\partial g_a(\xo)$.  The remaining $N-A$ genuine eigenvectors must be
orthogonal to $\partial g_a(\xo)$, and therefore span the tangent
subspace at $\xo$.  Their eigenvalues define the nature of $\xo$ as an
extremum of $V|_\mathcal{G}$.

The extremization problems considered in this paper involve a single
linear constraint with $\partial g(\xo) = (1,1,\ldots,1)$.  At a
two-component extremum $\xo$ of the form (\ref{eq:pot3c}), such
constraint leads to Hessian matrices of the form,
\begin{equation}
  \label{eq:popu3}
  \widetilde{H}_V(\xo) = \left(
    \begin{array}{c|c}
      \omega_3 I_{n_1}+ \kappa_1\dv{G}{n_1,n_1} &
      \kappa_2\dv{G}{n_1,n_2}\\\hline 
      \kappa_3\dv{G}{n_2,n_1} &
      \omega_4 I_{n_2}+\kappa_4\dv{G}{n_2,n_2} 
    \end{array}
    \right)~,
\end{equation}
with $\dv{G}{n_1,n_2}\in \mathbb{R}^{n_1\times n_2}$,
$(\dv{G}{n_1,n_2})_{ij}=1$, $I_n$ the $n\times n$ identity matrix, and
$\omega_{3,4}$, $\kappa_{1\cdots 4}$ some numerical coefficients.
Such matrices are easily diagonalized.  Let
$\{\dv{v}{n}_1,\ldots,\dv{v}{n}_n\}$ be an orthogonal basis for
$\mathbb{R}^n$ such that $\dv{v}{n}_1=(1,1,\ldots,1)$.  We define the
$n\times n$ orthogonal matrix $\dv{U}{n}$
\begin{equation}
  \label{eq:popu1}
  \dv{U}{n}_{ij} = \frac{1}{\sqrt{\dv{v}{n}_j\cdot \dv{v}{n}_j}}
  (\dv{v}{n}_j)_i
  \qquad
  \text{(no summation over $j$)}~.
\end{equation}
The matrix $\dv{G}{n_1,n_2}= \dv{v}{n_1}_1\otimes \dv{v}{n_2}_1$ then
satisfies $ \left({\dv{U}{n_1}}^\dagger
  \dv{G}{n_1,n_2}\dv{U}{n_2}\right)_{ij} = \sqrt{n_1 n_2}\delta_{i1}
\delta_{j1}$.  Let $U$ be the $N\times N$ orthogonal matrix
\begin{equation}
  \label{eq:popu4}
  U = \left(
    \begin{array}{c|c}
      \dv{U}{n_1} & 0\\\hline
            0     & \dv{U}{n_2}
    \end{array}
    \right)~.
\end{equation}
With this definition the matrix $U^\dagger \widetilde{H}_V(\xo) U$ is
block diagonal, with one block equal to $\omega_3 I_{n_1-1}$,
another equal to $\omega_4 I_{n_2-1}$, and a $2\times 2$ block $ \left(
  \begin{smallmatrix} \omega_3+\kappa_1 n_1 & \kappa_2 \sqrt{n_1 n_2} \\
    \kappa_3 \sqrt{n_1 n_2} & \omega_4+\kappa_4 n_2 \end{smallmatrix}
\right)$ with eigenvalues $\omega_{1,2}$, one of which should vanish
due to the projection. At an extremum of type (\ref{eq:pot3a}) the
form of $\widetilde{H}_V(\xo)$ is analogous to (\ref{eq:popu3}), with
9 blocks instead of 4, and the diagonalization procedure is the same
as above.

\section{Mass spectrum in $\sun{N}$}
\label{sec:appb}

In the main text and in appendix \ref{sec:hess} we considered the
extrema of $V_N(x)$ and its Hessian matrix.  From those we can
immediately obtain the extrema and mass spectrum in orbit space.  (The
latter being the orbifold $\sun{N}/\SUN{N} =
\mathbb{R}^N_c/\mathcal{P}_N$, with $\mathbb{R}^N_c = \{x\in
\mathbb{R}^N/\sum_{j=1}^N x_j=0\}$ and $\mathcal{P}_N$ the permutation
group of $N$ elements.)  In this appendix we extend the Hessian matrix
of $V_N(x)$ to $V_N(\phi)$ over $\sun{N}$. In order to do so we expand
\begin{equation}
  \label{eq:car}
  \phi = \sum_{i=1}^N \phi_{i} E^{(ii)} + \sum_{i<j=1}^N \rho_{ij}
  \frac{1}{\sqrt{2}} \left(E^{(ij)}+E^{(ji)}\right) + i \sum_{i<j=1}^N
  \eta_{ij} \frac{1}{\sqrt{2}} \left(E^{(ij)}-E^{(ji)}\right)
\end{equation}
with $\left(E^{(ij)}\right)_{mn}= \delta_{im} \delta_{jn}$, $1\leq
i,j,m,n\leq N$, and use $\{\phi_i\}_{i=1}^N \cup
\{\rho_{ij}\}_{i<j=1}^N \cup \{\eta_{ij}\}_{i<j=1}^N$ as our
coordinates, constrained by $\sum_{i=1}^N \phi_i
=0$.\footnote{Alternatively we could use $N-1$ unconstrained
  coordinates for the diagonal elements, by expanding in a basis with
  diagonal matrices $H^{(i)}$ satisfying $\tr H^{(i)}=0$, instead of
  $E^{(ii)}$, e.g., a Cartan--Weyl basis.}
The Hessian matrix of $V_N(\phi)$ at a diagonal matrix
$\varphi=\diag(x_1,\ldots,x_N)$ is easily computed if we disregard the
tracelessness constraint,
\begin{subequations}
  \label{eq:car2}  
\begin{align}
    \frac{\partial^2 V_N}{\partial \phi_l \partial
      \phi_k}(\varphi) &= \left(H_V(x_1,\ldots,x_N)\right)_{lk}~,
    \label{eq:car2a}\\ 
    \frac{\partial^2 V_N}{\partial\rho_{ij}\partial\rho_{kl}}(\varphi)
    &= \frac{\partial^2
      V_N}{\partial\eta_{ij}\partial\eta_{kl}}(\varphi)\propto
    \delta_{ik} \delta_{jl}~, \label{eq:car2b}\\
    \frac{\partial^2 V_N}{\partial \rho_{ij}\partial\phi_k}(\varphi)&= 
    \frac{\partial^2 V_N}{\partial \eta_{ij}\partial\phi_k}(\varphi)=
    \frac{\partial^2 V_N}{\partial
      \rho_{ij}\partial\eta_{kl}}(\varphi)=0  \label{eq:car2c}
\end{align}
\end{subequations}
where $H_V(x)$ is the Hessian matrix of $V_N(x)$.  The effect of
taking the constraint into account is to substitute
$\widetilde{H}_V(x)$ for $H_V(x)$ in (\ref{eq:car2}), as discussed in
appendix \ref{sec:hess}.

The matrix (\ref{eq:car2}) is block diagonal.  At an extremum of type
(\ref{eq:pot3c}) the eigenvalues of the block $\widetilde{H}_V(x)$ are
$\omega_{1,2,3,4}$ as given in (\ref{eq:masspec}).  The other two
diagonal blocks are already diagonal, as indicated in
(\ref{eq:car2b}).  Explicit calculation shows that the diagonal
entries in those blocks vanish if the index $(k,l)$ is such that
$\xo_k\neq \xo_l$ (see (\ref{eq:pot3c})).  If $(k,l)$ is such that
$\xo_k=\xo_l=\eta_1$ (resp.\ $\eta_2$) then $\partial^2
V/\partial\rho_{kl}\partial\rho_{kl} = \omega_3$ (resp. $\omega_4$).
Counting the number of such pairs of indices gives the multiplicity of
$\omega_{3,4}$ in each of the blocks $\partial^2
V/\partial\rho_{kl}\partial\rho_{kl}$ and $\partial^2
V/\partial\eta_{kl}\partial\eta_{kl}$.  Adding those multiplicities to
the reduced ones in (\ref{eq:masspec}) results in the total
multiplicities shown in that equation.


\begin{thebibliography}{9}
\bibitem{xli74}  L.-F.\  Li, Phys.\ Rev.\ \textbf{D 9} (1974), 1723.
%
\bibitem{mur83} T.\ Murphy, L.\ O'Raifeartaigh, Nucl.\ Phys.\
  \textbf{B 229} (1983), 509.
%
\bibitem{col93} S.\ Coleman, \emph{Aspects of Symmetry}, Cambridge
  U. Press, Cambridge, 1993.
%
\bibitem{che91} T.-P.\ Cheng, L.-F.\ Li, \emph{Gauge Theory of
    Elementary Particle Physics}, Oxford U. Press, Oxford, 1991. 
%
\bibitem{bou03} A.\ O.\ Bouzas, Int.\ J.\ Mod.\  \textbf{A 18} (2003),
  3695; ibid.\ \textbf{21} (2006), 1157.
\end{thebibliography}
\end{document}